\documentclass[aps,prl,preprint,groupedaddress,showpacs]{revtex4-1}

\usepackage{graphicx,float}
\usepackage[normalem]{ulem}
\usepackage{color}
\usepackage[euler]{textgreek}

\begin{document}

\title{A universal method for depositing patterned materials \textit{in-situ}}

\author{Yifan Chen$^1$$^{\S}$}
\author{Siu Fai Hung$^1$$^{\S}$}
\author{Wing Ki Lo$^1$}
\author{Yang Chen$^1$}
\author{Yang Shen$^1$}
\author{Kim Kafenda$^1$}
\author{Jia Su$^{2,}$$^{3}$}
\author{Kangwei Xia$^{1*}$}
\author{Sen Yang$^{1,}$$^{4*}$}

\affiliation{$^1$ Department of Physics, The Chinese University of Hong Kong, Shatin, New Territories, Hong Kong, China}
\affiliation{$^2$ Department of Biology, South University of Science and Technology of China, Shenzhen, Guangdong 518058, China}
\affiliation{$^3$ Shenzhen 34 Technology Co.,Ltd, Qianhai, Nanshan, Shenzhen, China, 518055}
\affiliation{$^4$ Shenzhen Research Institute, The Chinese University of Hong Kong, Shatin, New Territories, Hong Kong, China}
\affiliation{$^{\S}$ These authors contributed equally to this work}

\begin{abstract}
\end{abstract}

\maketitle

\textbf{Current techniques of patterned material deposition require separate steps for patterning and material deposition. The complexity and harsh working conditions post serious limitations for fabrication. Here, we introduce a novel single-step and easy-to-adapt method that can deposit materials \textit{in-situ}. Its unique methodology is based on the semiconductor nanoparticle assisted photon-induced chemical reduction and optical trapping. This universal mechanism can be used for depositing a large selection of materials including metals, insulators and magnets, with quality on par with current technologies. Patterning with several materials together with optical-diffraction-limited resolution accuracy can be achieved from macroscopic to microscopic scale. Furthermore, the setup is naturally compatible with optical microscopy based measurements, thus sample characterisation and material deposition can be realised \textit{in-situ}. Various devices fabricated with this method in 2D or 3D show it is ready for deployment in practical applications. This revolutionary method will provide a distinct tool in material technology.}

Conventional manufactory processes, ranging from circuit board printing down to integrated circuit and nano-devices fabrication, consist of multiple steps such as mask production, material deposition, photo-lithography and lift-off. Not only does each step add cost and chance to fail, it puts demanding requirements. For example, the sample surface temperature can be heated up too high during electron beam evaporation. Organic chemicals used in the photo-lithography process or high vacuum in deposition process can degrade the sample quality and put further limitations on the sample to use. Especially during lift-off, which is not trivial, faults lead to a waste of efforts and materials. And due to the complication of these procedures, advanced equipment and intensive training are required. 
On the other hand, precise positioning is a common difficulty under the current protocol, due to many factors, for example, the limited visibility of the sample surface covered with photoresists, moving samples among different setups in each procedure, imaging and aligning them without exposure. For the fabrications targeting on special sample locations, patterning while evaluating the sample quality is hard, as the process is not compatible with most characterisation measurements.

To invent a general patterned material deposition method with single step is desired to solve these problems. One of the key challenges is to find a mechanism to pattern the materials while depositing them \textit{in-situ}. The mechanism has to be general enough that it can be applied to commonly used materials, and the deposition should be big and thick enough in scale, and have quality suitable for real applications. There are a variety of ways for patterning polymer, including thermal process, photochemistry and optical trapping \cite{Kodama_81,Kawata_97, Ito_2011}. However, to pattern large scale inorganic materials directly, especially metals is still challenging. In general, up to now, only photon based methods can reach high resolution and accuracy. But different methods are based on different light induced effects. The methods based on photothermal induced physical form changes usually requires high power lasers, nevertheless fabrication with high resolution and with materials like noble metals in small scales are difficult. Laser induced chemical reactions, such as polymerization and reduction, can deposit nanostructures with selected materials \cite{Bjerneld_2002,Boyden_2018,Talapin_2017}. Optical trapping can also be used for depositing nanostructures either directly from their solution\cite{Miyasaka_2017,Zheng_2019,Park_2019}, or from photon-induced chemical reactions \cite{Itoh_2009}. However, without a bonding mechanism among particles, the maximum size and thickness, the mechanical strength and electrical conductance are greatly compromised in the finished structures. There are also approaches by mixing metal particle with polymer and using polymer printing techniques to build metal composite structures. But the conductivity is usually orders of magnitude smaller than the bulk value. Moreover, for the polymer based methods \cite{Talapin_2017,Ito_2011}, after the deposition, the development of samples with organic solvents is required, this posts further limitations on both materials to be deposited and the substrate to use. To solve all these problems, in this work, we present a novel single-step direct material deposition method based on the combination of the photon-induced chemical reduction and laser-induced optical trapping processes via semiconductor nanoparticles as the agent. The nanoparticles not only become a universal reduction agent and greatly broaden the selection of materials, but also can work as the seeds of growth trapped by the optical force as well as a "glue" to form rigid composite depositions. This new approach enables us to deposit a broad selection of materials with large area and thickness, high quality and high accuracy. It removes fabrication limitations posted by traditional methods and makes devices ready for practical applications.


\begin{figure*}
\center
\includegraphics[width=0.50\textwidth]{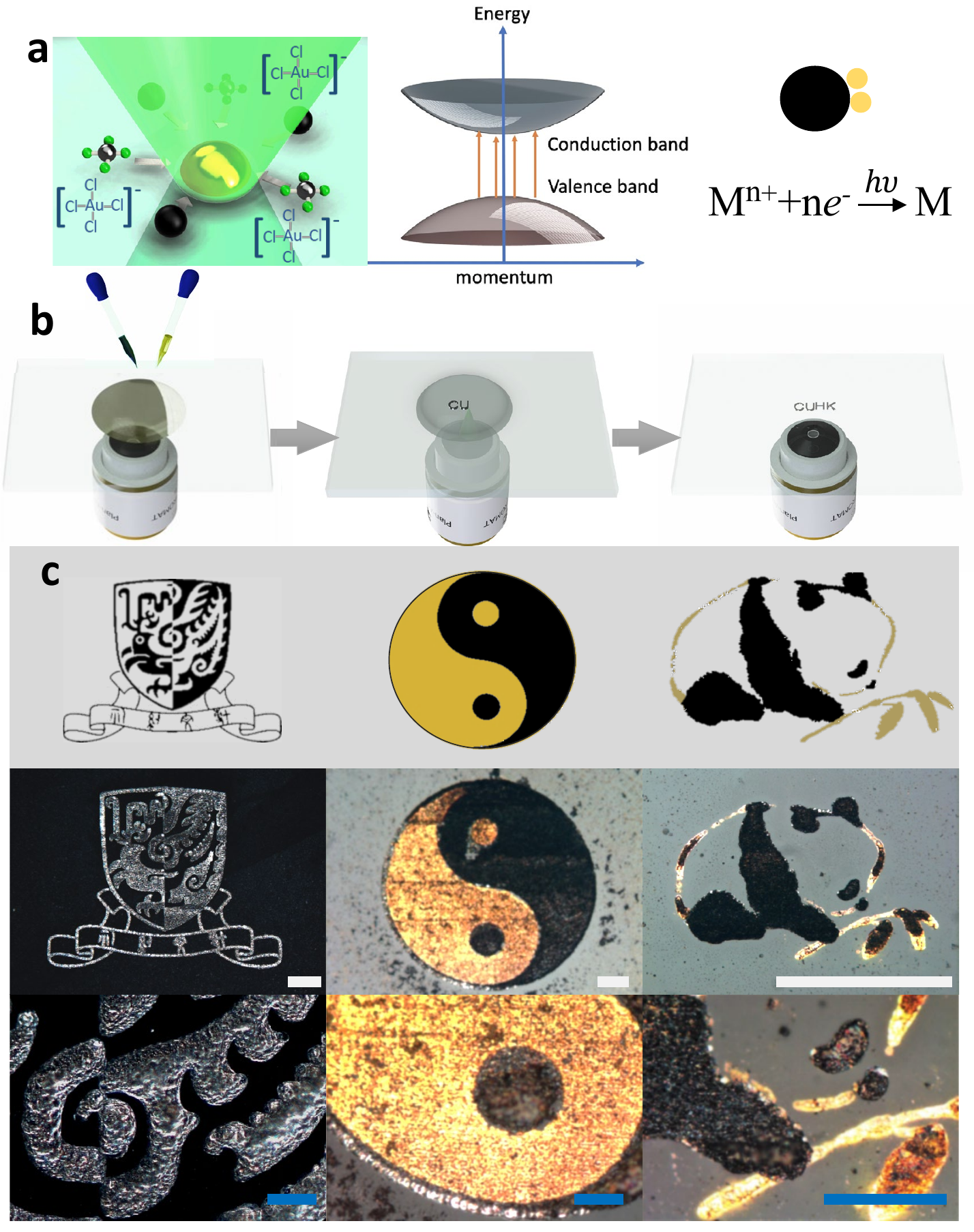}
	\caption{\textbf{Laser induced material deposition.}
   \textbf{a}, schematic illustrations of the principle of the LIMD method. (left) Schematic illustration of the material deposition process, i.e. under the optical field, the semiconductor nanoparticles are trapped towards the focus and on their surface photo-induced reduction reaction converts the metal ion into metal deposition. (middle) Ideal band structure of semiconductor particles. Photon excites electron from the valence band into the conduction band to become the free electron which triggers the photo-induced reduction reaction. (right) Schematic illustration of the photo-induced chemical reduction process. Free electrons excited by photon in the semiconductor particle (black sphere) reduce metal ions into metal particles (yellow sphere) on the surface of the semiconductor. \textbf{b}, schematic illustrations of the experimental procedure. (left) Drop-casted the thoroughly mixed reagents on the substrate surface, with one pipette contains metallate and the other contains semiconductor nanoparticles. (middle) a 532nm laser beam, focused by a microscope objective, creates deposition on the focal spot. (right) After the deposition and washing with water, the deposited pattern is left on the substrate. \textbf{c}, three samples of depositions on glass slides done with the LIMD method. The designs are shown in the upper row. The images in the lower row are the detail zoom-ins of the middle rows. (left) A logo of CUHK written in platinum. (centre) A yin-yang symbol written with platinum and gold. (right) A Chinese traditional ink painting written with platinum and gold. The white scale bars represent 50~\textmu m, the blue ones represent 25~\textmu m.}
\label{fig:1}
\end{figure*}

The working principle of the light induced material deposition (LIMD) is illustrated in the schematic drawings in Fig.~1(a). The reagent of this LIMD method consists of two water based solutions: Part A, contains mainly metallate; Part B, contains semiconductor nanoparticles. When a continuous wave (CW) laser is focused on the reagent, free electrons, excited from the valence band of the semiconductor nanoparticle by photons, trigger a chemical reduction process, which converts metal ions in the solution to metal particles on the surface of the semiconductor nanoparticle. Simultaneously, the focused laser beam also works as an optical trap driving particles towards the focus spot near the substrate surface \cite{Ashkin_70,Ashkin_86}. Unlike commonly used water-solvable reduction agents, these semiconductor nanoparticles can couple these two processes together. Thus, the chemically-reduced metal growing on the surface of the particles works like a glue, bonding trapped particles together to form a mechanically rigid metal/semiconductor composite on the substrate surface (discussions on mechanism in Supplementary Information).


The whole fabrication process is illustrated in Fig.~1(b). Part A (the metallate) and Part B (the semiconductor nanoparticles) are chosen based on the material to be deposited. Thoroughly mixed solution of these two is drop-casted on the substrate surface. A home-built laser writing system is used to write the patterned material \textit{in-situ}. While keeping the sample in the setup, the residue reagent is sucked away, then the deposited material and substrate are cleaned with water. Examples from macroscopic scale down to microscopic scale of this LIMD method and their zoom-in details are shown in Fig.~1(c) (experimental details in Supplementary Information). On the left, we wrote the logo of The Chinese University of Hong Kong (CUHK) in platinum on a glass substrate. Making a similarly complicated structure in small scale using photo-lithography is difficult as either incorrect exposures or lift-off failure can ruin the fine details in the pattern. Since there are no development or lift-off steps, this complicated pattern came out straightaway.

Depositing patterns with two or more different materials, an important task in nanotechnology, is even more difficult for traditional fabrication methods, because multiple rounds of material depositions and photo-lithography with fine and accurate alignments are required. It is even harder to reach a good alignment among multiple rounds of material deposition within the whole area. However, it becomes a simple task with the LIMD method. To deposit different materials, above procedure just has to be repeated. In between two depositions, the reagent is changed without removing the sample from the setup. Thus, the alignment can be kept within the optical resolution of 400~nm. Here, as an example, in the centre of Fig.~1(c), a Yin-Yang-fish symbol was written with gold and platinum. On the right of Fig.~1(c), we drew a Chinese traditional ink painting with platinum and gold: the dark body parts of the panda were drawn in platinum, and the rest, including the contour of the panda and the bamboo, were drawn in gold. In both examples, with the LIMD method, these patterns were written with the least effort needed but high fidelity and submicron precision. Furthermore, by employing microfluidic chips with multiple channels in the future \cite{Whitesides2006}, depositions with different materials can be simplified further and automated.

Owing to its universality in its mechanism, this method can be applied to a vast selection of materials and substrates. Based on the material to be deposited, the corresponding Part A component can be chosen. In Fig.~2, we show the collection of combinations we have tested. To be noted, this method works well with the commonly used noble metals, including gold, silver and platinum, as well as the widely used substrates, such as glass, quartz, sapphire and indium tin oxide (ITO) (see Supplementary Information). It can also be used for depositing transition metals and substrates like flexible membranes and even tapes. This vast material possibility shows the potential of this LIMD method in general applications.

\begin{figure*}
\includegraphics[width=1.0\textwidth]{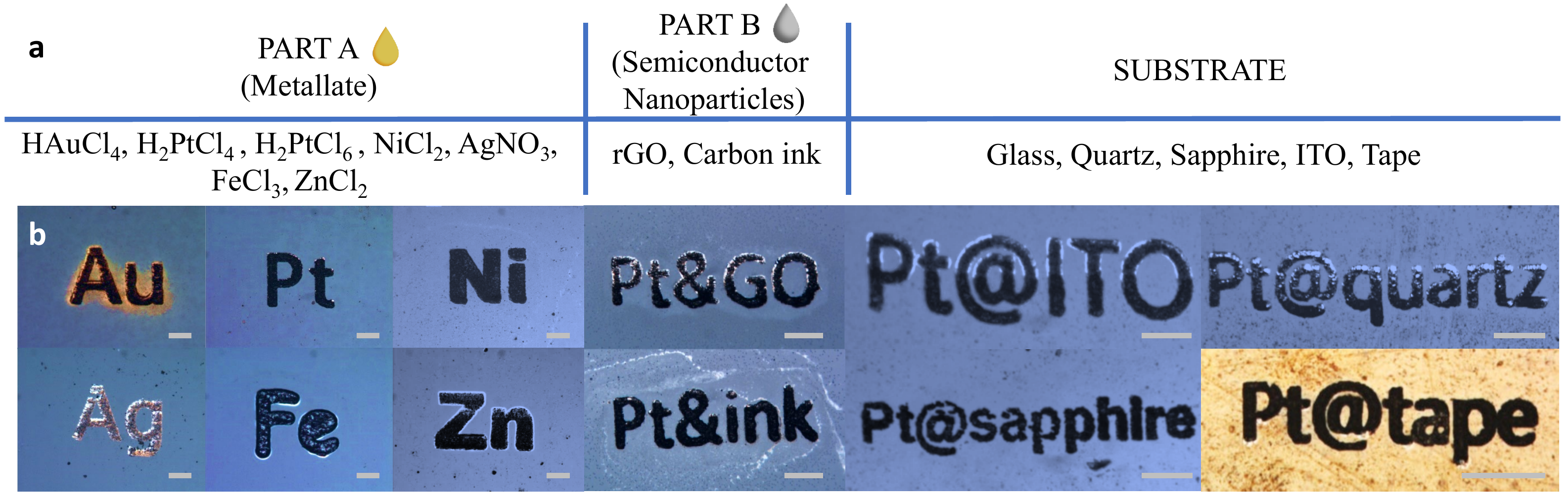}
	\caption{\textbf{Material depositions with the LIMD method on various substrates.}
   \textbf{a}, list of the ingredients and substrates used by the LIMD method in this work. \textbf{b}, reflection type illumination microscopic images of different materials deposited on substrates. Except for the fourth one from the left in the upper row, which was done with reduced graphene oxide as Part B, the rest were done with carbon ink. Energy-dispersive X-ray (EDX) spectroscopy data which show the element compositions in the layer are available in Supplementary Information. The grey scale bars represent 50~\textmu m.}
\label{fig:2}
\end{figure*}

Part B in the recipe works as the reducing agent for the chemical reduction process. In general, all nanoparticles with a small band gap can work. Carbon based nanoparticles, for example, are a good choice. Since the particles are required to mix with water based metallate solution, it is better to have them dispersed uniformly in water. Reduced graphene oxide fulfils these requirements and performs well. However, to our big surprise, the best performance was achieved using off-the-shelf carbon ink originally designed for fountain pens and Chinese ink brushes. After generations of optimisations by commercial companies, these ink particles not only are dispersed uniformly inside water, but have nearly identical hydrodynamic diameters down to 100~nm (see Supplementary Information). This feature gives the LIMD method even more advantages compared with traditional methods, such as cheap, easy to adapt, and green to the environment.

The quality of the deposition by the LIMD method can be evaluated from its physical structure and its physical performance. To estimate the spatial resolution reached by the LIMD method, both nano-dots and nano-wire are deposited by using high N.A. oil objective lens and fine tuned parameters. The scanning electron microscope (SEM) images of platinum dots with size around 0.5~\textmu m, and that of a 1~\textmu m wide iron oxide wire are shown in Fig.~3(a,b) respectively. Also the cross-section of the iron oxide wire, cut with a focused ion beam (FIB) machine, is shown in Fig.~3(b). The deposition is uniform in size and solid inside. The surface roughness was around 30~nm, as measured by an atomic force microscope (see Supplementary Information).

\begin{figure*}
\center
\includegraphics[width=0.95\textwidth]{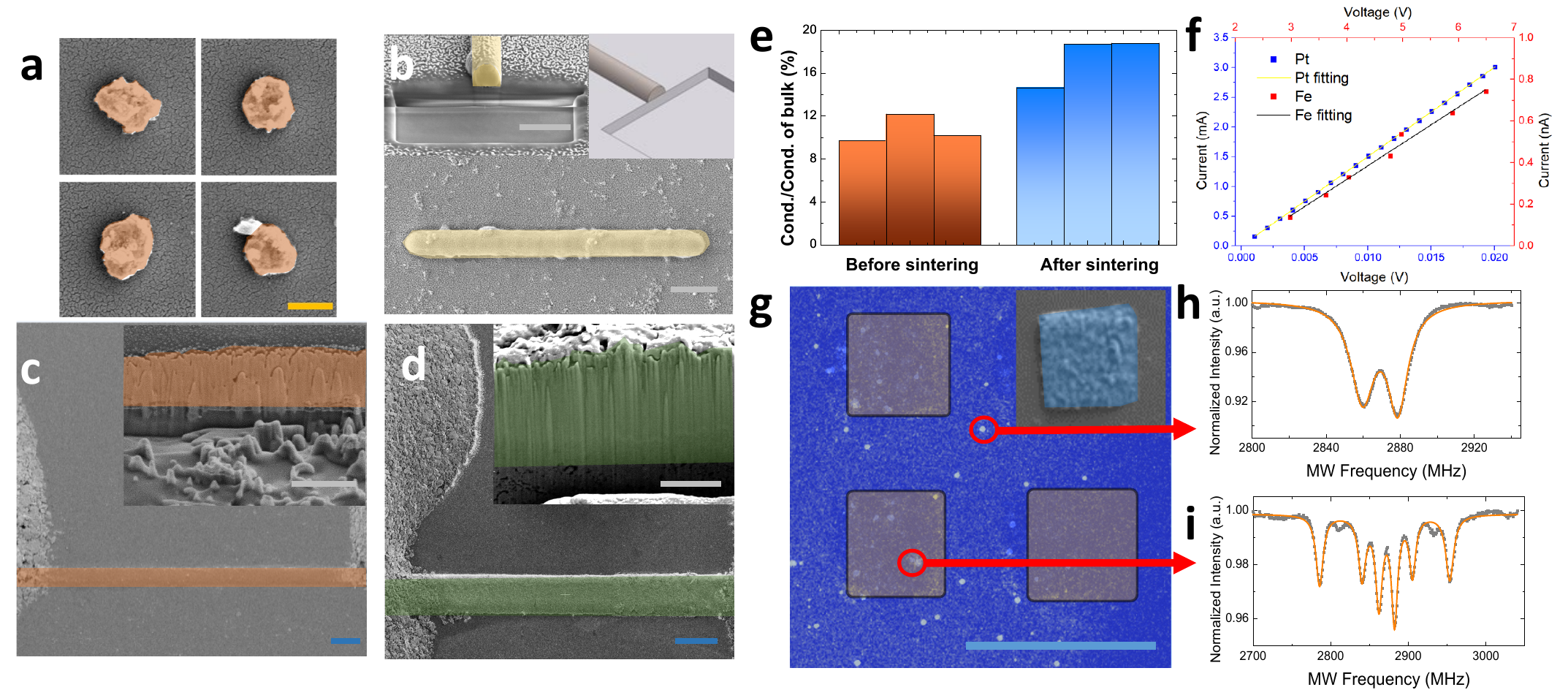}
	\caption{\textbf{Physical properties of the deposited structures.}
   \textbf{a, b}, SEM images show the spatial resolution reached by the LIMD method in depositing platinum dots (a) and iron oxide nano-wire (b) on glass slides. The schematic drawing of the deposition and the SEM image of cross-section view are shown as the inserts in (b). \textbf{c, d}, SEM images of the surface view and cross-section view (inserts) of platinum deposition (c) and iron oxide deposition (d) on glass slides. \textbf{e}, the conductivity of platinum structures made before (orange) and after (blue) laser sintering. \textbf{f}, the \textit{I-V} curves measured for platinum (blue) and iron oxide (red) samples. \textbf{g}, the fluorescence image from the confocal microscopy scan and the SEM image (insert) of nickel depositions on a glass slide. \textbf{h, i}, the ESR spectra measured with NV centres in diamond in the middle point of the four nickel pads (h) and on top of one Nickel pad (i). The grey scale bars represent 2~\textmu m, the blue ones represent 50~\textmu m, the yellow one represents 500~nm. Experimental details are in Supplementary Information.}
   \label{fig:3}
\end{figure*}

Noble metals are widely used as electrodes and wires in real applications due to their excellent conductance. Thus, we performed electrical conductivity measurements (results shown in Fig.~3(e)), on a platinum wire deposited on a glass substrate (SEM image in Fig.~3(c)). After the LIMD deposition, the conductivity is around $10\%$ of the bulk value of platinum. To examine the origin of this relatively low conductivity, the SEM image of its cross-section is shown in the insert in Fig.~3(c). Unlike the cases in Fig.~3(a,b), to reach the deposition in macroscopic scale in high writing speed here, low N.A. air objective lens and higher laser power are used. Under these conditions, the deposition consists of a stacking of pillars. This structure may come from the Turing instability \cite{Turing}. The typical diameter of the pillar structure, which is around a few hundred nanometres, is mainly determined by the penetration depth of the light in the materials (see Supplementary Information). Also, since this reaction-diffusion instability mechanism tends to deplete the resources in between reaction centres, the connections among pillar surfaces can be weak. This leads to the low conductivity. To solve this problem, the structure was sintered in air in the setup with the same laser \cite{Boyden_2018}. The resulting conductivity rises to $18\%$ of the bulk value (detail analysis in Supplementary Information).

With this LIMD method, not only conductors, but also insulators can be deposited. One approach is to use iron based metallate as Part A. Iron is so chemically active that even by using oxygen-reduced solution, the deposition ends up as iron oxide instead of pure iron. It turns out to be a good insulator as shown in the I-V curve measurement in Fig.~3(f). The SEM image of the cross-section in Fig.~3(d) proves the structure is compact-packed. Therefore, this iron oxide deposition can be used as a practical insulator for general purpose.

Besides electrical conductance, we also calibrate their mechanical properties. By using nanoindentation method, the Young's Modulus of Pt deposition is determined to be 1.0~GPa. The number is similar to graphite which is the softer component in the deposition. On the other hand, the deposition is elastic and flexible due to this composite nature (detail analysis in Supplementary Information). This kind of high conductance and high flexibility composite material is unique and ideal for making flexible electronics.

The LIMD method can be used to deposit ferromagnetic materials as well. Take nickel as an example. We wrote four square shaped Ni layers with a permanent magnet polarizing them during the deposition process, as shown in Fig.~3(g). The magnetic profile then was measured with nitrogen vacancy (NV) centres inside nanodiamonds spread on the sample \cite{Wrachtrup06,Wrachtrup08,Lukin08,Yang_2019}. The splitting of electron spin resonance (ESR) lines from these NV centres, shown in Fig.~3(h,i), is due to the Zeeman effect of the residue magnetic field from the Ni layers. The field is nearly 30~Gauss inside the Ni square while close to none outside (details in Supplementary Information). This implies the ferromagnetic property in the deposited layer. Thus, this LIMD method enables us to \textit{in-situ} fabricate micromagnets and complicated magnetic structures.

\begin{figure*}
\center
\includegraphics[width=1.0\textwidth]{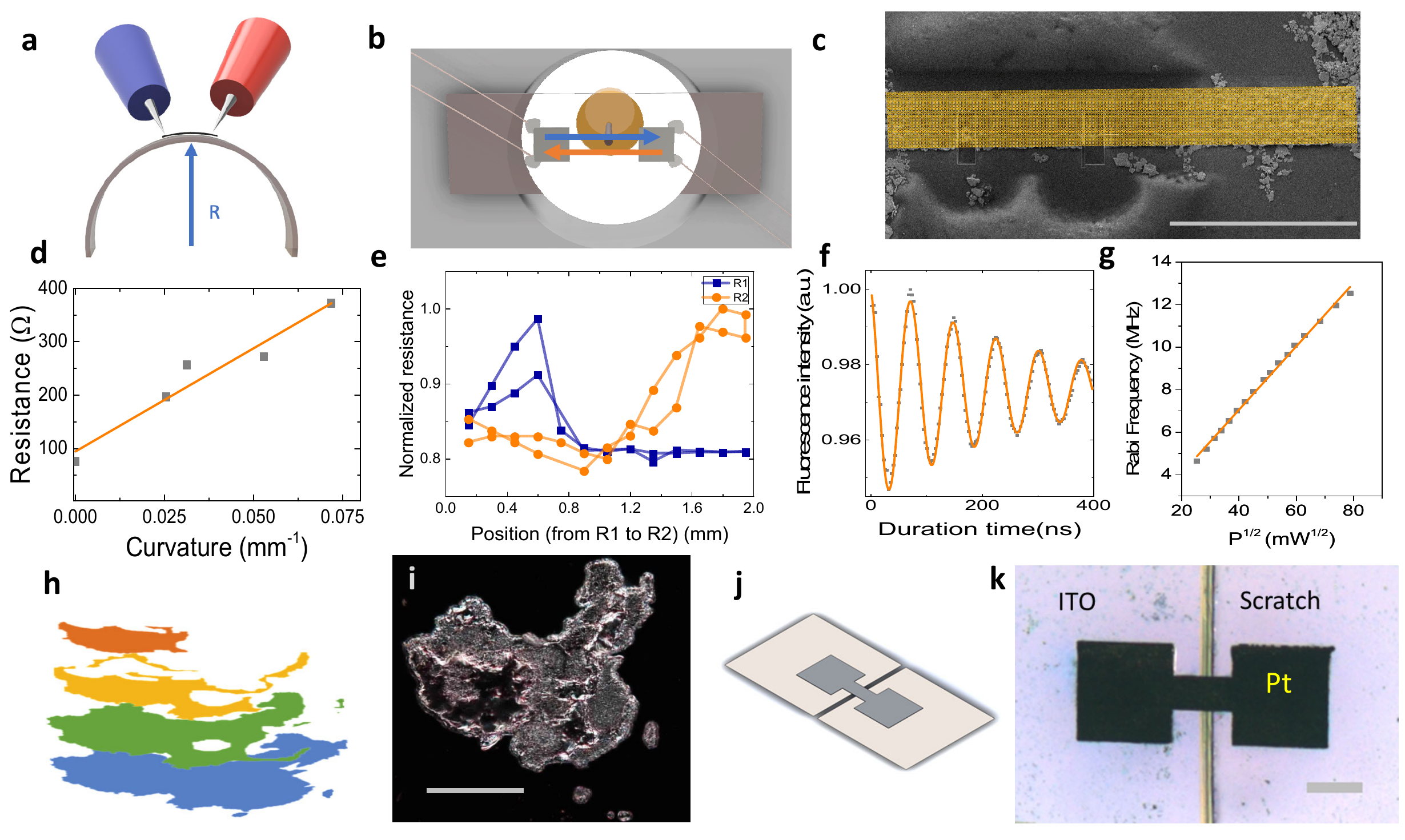}
	\caption{\textbf{Devices made by the LIMD method for practical applications.}
   \textbf{a-b}, schematic illustrations of two flexible devices: a resistive flex sensor (a) and a resistance-based touch sensor (b) (more details in Supplementary Information). \textbf{c}, the SEM image of the microwave waveguide coupled with nanodiamond particles. \textbf{d}, resistance dependence with curvature measured in the device in (a). \textbf{e}, resistance dependence of two platinum squares in the device (b) with position of touching. The resistance is normalized to the values without touching. \textbf{f}, the microwave driven Rabi oscillation of electron spins in NV centres in diamond in the vicinity of the waveguide in (c). \textbf{g}, the microwave power dependence of the Rabi oscillation frequency. The microwave power was measured after the microwave power amplifier before entering the microwave waveguide. \textbf{h}, design patterns for 3D laser writing. The structure was written in the layer by layer manner. The raw topography data were obtained from the Ministry of Natural Resources of China. \textbf{i}, a 3D topographic map made by the iron oxide deposition. \textbf{j, k}, the schematic design (j) and the microscopic image (k) of repairing the gap between ITO contacts with the LIMD method. The grey scale bars represent 100~\textmu m.}
\label{fig:4}
\end{figure*}

The performance of the deposited materials can be further demonstrated in three application devices. Flexible electronics and wearable technology are booming fields \cite{Gates,Kim2019}. The commonly used flexible substrates usually have low melting points. Traditional deposition methods such as metal thermal evaporation may cause melting of the sample surface. This LIMD method has a low working temperature (see Supplementary Information) and is water-solution based. It even works well with various thin tapes. Furthermore, unlike other metal contacts, this composite metal deposition has a conductance similar to metal while being flexible like a polymer. On the other hand, simplification in production and flexibility in design can be key factors for commercial developments. The direct laser writing nature of the LIMD method provides a simple and cheap way for production and customisation. As examples, we build two proof-of-principle devices: a resistive flex sensor and a resistive touch sensor. Resistive flex sensors are important parts for robotics \cite{Saggio_2015}. The schematic drawing of this sensor is shown in Fig.~4(a). A platinum line with length of 300~\textmu m and width of 50~\textmu m was written on a Kapton tape. The resistance of the wire shows a linear dependency with the curvature, as shown in Fig.~4(d). The device reaches similar sensitivity of current sensors, but is two orders of magnitude smaller in size \cite{Saggio_2015}.

From this deformation-sensitive mechanism, resistive touch sensor devices can be further developed \cite{Walker_12}. As shown in Fig.~4(b), two parallel platinum squares were written on a Kapton tape. Depending on the location where pressure was applied on, the resistances of both lines show different trend of change, as shown in Fig.~4(e). Based on this dependence, a measurement of both resistances can determine the location of the touching.

Besides applications in macroscopic scale, with this method also devices for microscopic applications can be fabricated with submicron accuracy. One of the major road blocks for nanotechnology is manufacturing high performance devices with high precision with respect to tiny or chosen samples \cite{Awschalom_18}. The LIMD method provides a unique solution as the deposition and imaging are \textit{in-situ}. One key device in solid state based quantum information science is the microwave waveguide, which is used to enhance coupling of microwave radiation to solid state qubits \cite{Wrachtrup06,Fuchs_09}. This waveguide has to be in close vicinity to the tiny qubits and it should be able to transmit high power microwave signals. Shown in Fig.~4(c), combining both confocal microscopy imaging and the LIMD system, we directly wrote a microwave waveguide near a nanodiamond particle. This waveguide performs as good as structures made with conventional method, as shown in the electron spin Rabi oscillation measurement (Fig.~4(g)). With a cross-section of only $27\times2$~\textmu m$^2$, this structure works well for a large power range. Shown in Fig.~4(g), is the linear dependency of the Rabi frequency to the square root of the applied microwave power up to 6.2~Watts. Furthermore, since the laser writing setup is compatible with other microscopy-based setups, it is possible to combine both the sample characterisation measurements and the LIMD method together. For example, in Supplementary Information, while depositing materials on a nanodiamond with the LIMD method, we simultaneously use the NV centres inside the nanodiamond as a thermometer to measure a possible heating effect. Methods like this will greatly benefit the research fields where a big variation of the sample properties are unavoidable, such as two dimensional materials, nanoparticles and nanostructures \cite{Yao_16,Bao_2013,High2015}.

With this LIMD method, not only 2D, but also 3D patterns can be made, due to its unique mechanisms. By moving the focus in vertical direction, 3D material deposition can be done in the layer by layer manner. As an example, a 3D topographic map was written and shown in Fig.~4(i). Unlike other 3D laser writing methods with the requirement of high power ultrafast laser, a CW laser with $\sim100$~mW is enough for the LIMD method. This makes this method one of the simplest methods for 3D laser writing, as well as a unique method which is practical in both 2D and 3D fabrication.

Reflow soldering is one of the most widely used bonding methods. We further demonstrated the reflow soldering working with the structures made by the LIMD method (see Supplementary Information). This paves the way for its application in modern IC industry.

One more special application of this LIMD method is repairing circuits. After the circuit has been fabricated, it becomes hard to repair a small crack or a broken pad with photolithography method. It is tricky to use the conductive epoxy to reach high spatial resolution. This LIMD method can provide an unique solution. We demonstrate this with an indium tin oxide (ITO) structure commonly used for solar cell. A 20~\textmu m crack was cut by a diamond cutter in the middle of the ITO pad. As shown in Fig.~4 (j,k), a Pt bridge is made via the LIMD method to restore the connection (see Supplementary Information for details).


It is worth to mention that, compared to other conventional microstructure fabrication devices, this microscopy based LIMD setup also reduces the requirements on equipment and training dramatically. An ink-jet laser printer for directly printing materials could even be made with this method. It would make the print-out of electrical circuits in the field possible, and greatly reduce the fabrication costs in labour, material and equipment. After the deposition with the LIMD method, all the residue solution can be recycled to minimize material waste and pollution. Moreover, unlike photo-lithography method where the organic solvent often puts the sample surface in risk, there is no organic photoresist or solvent involved in this method. As current electrical fabrications produce enormous amounts of waste, including organic solvents, chemical etching wastes and left-over materials, which endanger the whole environment, this nearly waste free method will play a big role in reforming the current industry towards a greener one.

In summary, we introduce a novel patterned material deposition method based on the combination of photon-induced chemical reduction and laser-induced optical trapping with semiconductor nanoparticles as the agents. 
In this first demonstration, we already showed the successful applications with a vast selection of materials on a large selection of substrates. Much more recipes can be designed based on this general mechanism. For example, the nanoparticles can be inorganic semiconductor or perovskite structures, or other organic materials. With more experts from different disciplines joining, the LIMD method will be developed into a powerful tool in the modern material research.

\bibliography{references}

\begin{thebibliography}{27}%
\makeatletter
\providecommand \@ifxundefined [1]{%
 \@ifx{#1\undefined}
}%
\providecommand \@ifnum [1]{%
 \ifnum #1\expandafter \@firstoftwo
 \else \expandafter \@secondoftwo
 \fi
}%
\providecommand \@ifx [1]{%
 \ifx #1\expandafter \@firstoftwo
 \else \expandafter \@secondoftwo
 \fi
}%
\providecommand \natexlab [1]{#1}%
\providecommand \enquote  [1]{``#1''}%
\providecommand \bibnamefont  [1]{#1}%
\providecommand \bibfnamefont [1]{#1}%
\providecommand \citenamefont [1]{#1}%
\providecommand \href@noop [0]{\@secondoftwo}%
\providecommand \href [0]{\begingroup \@sanitize@url \@href}%
\providecommand \@href[1]{\@@startlink{#1}\@@href}%
\providecommand \@@href[1]{\endgroup#1\@@endlink}%
\providecommand \@sanitize@url [0]{\catcode `\\12\catcode `\$12\catcode
  `\&12\catcode `\#12\catcode `\^12\catcode `\_12\catcode `\%12\relax}%
\providecommand \@@startlink[1]{}%
\providecommand \@@endlink[0]{}%
\providecommand \url  [0]{\begingroup\@sanitize@url \@url }%
\providecommand \@url [1]{\endgroup\@href {#1}{\urlprefix }}%
\providecommand \urlprefix  [0]{URL }%
\providecommand \Eprint [0]{\href }%
\providecommand \doibase [0]{http://dx.doi.org/}%
\providecommand \selectlanguage [0]{\@gobble}%
\providecommand \bibinfo  [0]{\@secondoftwo}%
\providecommand \bibfield  [0]{\@secondoftwo}%
\providecommand \translation [1]{[#1]}%
\providecommand \BibitemOpen [0]{}%
\providecommand \bibitemStop [0]{}%
\providecommand \bibitemNoStop [0]{.\EOS\space}%
\providecommand \EOS [0]{\spacefactor3000\relax}%
\providecommand \BibitemShut  [1]{\csname bibitem#1\endcsname}%
\let\auto@bib@innerbib\@empty
\bibitem [{\citenamefont {Kodama}(1981)}]{Kodama_81}%
  \BibitemOpen
  \bibfield  {author} {\bibinfo {author} {\bibfnamefont {H.}~\bibnamefont
  {Kodama}},\ }\href {\doibase 10.1063/1.1136492} {\bibfield  {journal}
  {\bibinfo  {journal} {Review of Scientific Instruments}\ }\textbf {\bibinfo
  {volume} {52}},\ \bibinfo {pages} {1770} (\bibinfo {year}
  {1981})}\BibitemShut {NoStop}%
\bibitem [{\citenamefont {Shoji}\ \emph {et~al.}(1997)\citenamefont {Shoji},
  \citenamefont {Osamu},\ and\ \citenamefont {Satoshi}}]{Kawata_97}%
  \BibitemOpen
  \bibfield  {author} {\bibinfo {author} {\bibfnamefont {M.}~\bibnamefont
  {Shoji}}, \bibinfo {author} {\bibfnamefont {N.}~\bibnamefont {Osamu}}, \ and\
  \bibinfo {author} {\bibfnamefont {K.}~\bibnamefont {Satoshi}},\ }\href@noop
  {} {\bibfield  {journal} {\bibinfo  {journal} {Opt. Lett.}\ }\textbf
  {\bibinfo {volume} {22}},\ \bibinfo {pages} {132} (\bibinfo {year}
  {1997})}\BibitemShut {NoStop}%
\bibitem [{\citenamefont {Ito}\ \emph {et~al.}(2011)\citenamefont {Ito},
  \citenamefont {Tanaka}, \citenamefont {Yoshikawa}, \citenamefont {Ishibashi},
  \citenamefont {Miyasaka},\ and\ \citenamefont {Masuhara}}]{Ito_2011}%
  \BibitemOpen
  \bibfield  {author} {\bibinfo {author} {\bibfnamefont {S.}~\bibnamefont
  {Ito}}, \bibinfo {author} {\bibfnamefont {Y.}~\bibnamefont {Tanaka}},
  \bibinfo {author} {\bibfnamefont {H.}~\bibnamefont {Yoshikawa}}, \bibinfo
  {author} {\bibfnamefont {Y.}~\bibnamefont {Ishibashi}}, \bibinfo {author}
  {\bibfnamefont {H.}~\bibnamefont {Miyasaka}}, \ and\ \bibinfo {author}
  {\bibfnamefont {H.}~\bibnamefont {Masuhara}},\ }\href@noop {} {\bibfield
  {journal} {\bibinfo  {journal} {Journal of the American Chemical Society}\
  }\textbf {\bibinfo {volume} {133}},\ \bibinfo {pages} {14472} (\bibinfo
  {year} {2011})}\BibitemShut {NoStop}%
\bibitem [{\citenamefont {Bjerneld}\ \emph {et~al.}(2002)\citenamefont
  {Bjerneld}, \citenamefont {Murty}, \citenamefont {Prikulis},\ and\
  \citenamefont {Käll}}]{Bjerneld_2002}%
  \BibitemOpen
  \bibfield  {author} {\bibinfo {author} {\bibfnamefont {E.~J.}\ \bibnamefont
  {Bjerneld}}, \bibinfo {author} {\bibfnamefont {K.~V. G.~K.}\ \bibnamefont
  {Murty}}, \bibinfo {author} {\bibfnamefont {J.}~\bibnamefont {Prikulis}}, \
  and\ \bibinfo {author} {\bibfnamefont {M.}~\bibnamefont {Käll}},\
  }\href@noop {} {\bibfield  {journal} {\bibinfo  {journal} {ChemPhysChem}\
  }\textbf {\bibinfo {volume} {3}},\ \bibinfo {pages} {116} (\bibinfo {year}
  {2002})}\BibitemShut {NoStop}%
\bibitem [{\citenamefont {Oran}\ \emph {et~al.}(2018)\citenamefont {Oran},
  \citenamefont {Rodriques}, \citenamefont {Gao}, \citenamefont {Asano},
  \citenamefont {Skylar-Scott}, \citenamefont {Chen}, \citenamefont {Tillberg},
  \citenamefont {Marblestone},\ and\ \citenamefont {Boyden}}]{Boyden_2018}%
  \BibitemOpen
  \bibfield  {author} {\bibinfo {author} {\bibfnamefont {D.}~\bibnamefont
  {Oran}}, \bibinfo {author} {\bibfnamefont {S.~G.}\ \bibnamefont {Rodriques}},
  \bibinfo {author} {\bibfnamefont {R.}~\bibnamefont {Gao}}, \bibinfo {author}
  {\bibfnamefont {S.}~\bibnamefont {Asano}}, \bibinfo {author} {\bibfnamefont
  {M.~A.}\ \bibnamefont {Skylar-Scott}}, \bibinfo {author} {\bibfnamefont
  {F.}~\bibnamefont {Chen}}, \bibinfo {author} {\bibfnamefont {P.~W.}\
  \bibnamefont {Tillberg}}, \bibinfo {author} {\bibfnamefont {A.~H.}\
  \bibnamefont {Marblestone}}, \ and\ \bibinfo {author} {\bibfnamefont {E.~S.}\
  \bibnamefont {Boyden}},\ }\href {\doibase 10.1126/science.aau5119} {\bibfield
   {journal} {\bibinfo  {journal} {Science}\ }\textbf {\bibinfo {volume}
  {362}},\ \bibinfo {pages} {1281} (\bibinfo {year} {2018})}\BibitemShut
  {NoStop}%
\bibitem [{\citenamefont {Wang}\ \emph {et~al.}(2017)\citenamefont {Wang},
  \citenamefont {Fedin}, \citenamefont {Zhang},\ and\ \citenamefont
  {Talapin}}]{Talapin_2017}%
  \BibitemOpen
  \bibfield  {author} {\bibinfo {author} {\bibfnamefont {Y.}~\bibnamefont
  {Wang}}, \bibinfo {author} {\bibfnamefont {I.}~\bibnamefont {Fedin}},
  \bibinfo {author} {\bibfnamefont {H.}~\bibnamefont {Zhang}}, \ and\ \bibinfo
  {author} {\bibfnamefont {D.~V.}\ \bibnamefont {Talapin}},\ }\href {\doibase
  10.1126/science.aan2958} {\bibfield  {journal} {\bibinfo  {journal}
  {Science}\ }\textbf {\bibinfo {volume} {357}},\ \bibinfo {pages} {385}
  (\bibinfo {year} {2017})}\BibitemShut {NoStop}%
\bibitem [{\citenamefont {Setoura}\ \emph {et~al.}(2017)\citenamefont
  {Setoura}, \citenamefont {Ito}, \citenamefont {Yamada}, \citenamefont
  {Yamauchi},\ and\ \citenamefont {Miyasaka}}]{Miyasaka_2017}%
  \BibitemOpen
  \bibfield  {author} {\bibinfo {author} {\bibfnamefont {K.}~\bibnamefont
  {Setoura}}, \bibinfo {author} {\bibfnamefont {S.}~\bibnamefont {Ito}},
  \bibinfo {author} {\bibfnamefont {M.}~\bibnamefont {Yamada}}, \bibinfo
  {author} {\bibfnamefont {H.}~\bibnamefont {Yamauchi}}, \ and\ \bibinfo
  {author} {\bibfnamefont {H.}~\bibnamefont {Miyasaka}},\ }\href {\doibase
  https://doi.org/10.1016/j.jphotochem.2017.05.002} {\bibfield  {journal}
  {\bibinfo  {journal} {Journal of Photochemistry and Photobiology A:
  Chemistry}\ }\textbf {\bibinfo {volume} {344}},\ \bibinfo {pages} {168 }
  (\bibinfo {year} {2017})}\BibitemShut {NoStop}%
\bibitem [{\citenamefont {Li}\ \emph {et~al.}(2019)\citenamefont {Li},
  \citenamefont {Hill}, \citenamefont {Lin},\ and\ \citenamefont
  {Zheng}}]{Zheng_2019}%
  \BibitemOpen
  \bibfield  {author} {\bibinfo {author} {\bibfnamefont {J.}~\bibnamefont
  {Li}}, \bibinfo {author} {\bibfnamefont {E.~H.}\ \bibnamefont {Hill}},
  \bibinfo {author} {\bibfnamefont {L.}~\bibnamefont {Lin}}, \ and\ \bibinfo
  {author} {\bibfnamefont {Y.}~\bibnamefont {Zheng}},\ }\href {\doibase
  10.1021/acsnano.9b01034} {\bibfield  {journal} {\bibinfo  {journal} {ACS
  Nano}\ }\textbf {\bibinfo {volume} {13}},\ \bibinfo {pages} {3783} (\bibinfo
  {year} {2019})}\BibitemShut {NoStop}%
\bibitem [{\citenamefont {Park}\ \emph {et~al.}(2019)\citenamefont {Park},
  \citenamefont {Kang}, \citenamefont {Baek}, \citenamefont {Park},
  \citenamefont {Oh}, \citenamefont {Lee}, \citenamefont {Koo},\ and\
  \citenamefont {Park}}]{Park_2019}%
  \BibitemOpen
  \bibfield  {author} {\bibinfo {author} {\bibfnamefont {J.-E.}\ \bibnamefont
  {Park}}, \bibinfo {author} {\bibfnamefont {H.~S.}\ \bibnamefont {Kang}},
  \bibinfo {author} {\bibfnamefont {J.}~\bibnamefont {Baek}}, \bibinfo {author}
  {\bibfnamefont {T.~H.}\ \bibnamefont {Park}}, \bibinfo {author}
  {\bibfnamefont {S.}~\bibnamefont {Oh}}, \bibinfo {author} {\bibfnamefont
  {H.}~\bibnamefont {Lee}}, \bibinfo {author} {\bibfnamefont {M.}~\bibnamefont
  {Koo}}, \ and\ \bibinfo {author} {\bibfnamefont {C.}~\bibnamefont {Park}},\
  }\href {\doibase 10.1021/acsnano.9b03405} {\bibfield  {journal} {\bibinfo
  {journal} {ACS Nano}\ }\textbf {\bibinfo {volume} {13}},\ \bibinfo {pages}
  {9122} (\bibinfo {year} {2019})}\BibitemShut {NoStop}%
\bibitem [{\citenamefont {Itoh}\ \emph {et~al.}(2009)\citenamefont {Itoh},
  \citenamefont {Biju}, \citenamefont {Ishikawa}, \citenamefont {Ito},\ and\
  \citenamefont {Miyasaka}}]{Itoh_2009}%
  \BibitemOpen
  \bibfield  {author} {\bibinfo {author} {\bibfnamefont {T.}~\bibnamefont
  {Itoh}}, \bibinfo {author} {\bibfnamefont {V.}~\bibnamefont {Biju}}, \bibinfo
  {author} {\bibfnamefont {M.}~\bibnamefont {Ishikawa}}, \bibinfo {author}
  {\bibfnamefont {S.}~\bibnamefont {Ito}}, \ and\ \bibinfo {author}
  {\bibfnamefont {H.}~\bibnamefont {Miyasaka}},\ }\href@noop {} {\bibfield
  {journal} {\bibinfo  {journal} {Applied Physics Letters}\ }\textbf {\bibinfo
  {volume} {94}},\ \bibinfo {pages} {144105} (\bibinfo {year}
  {2009})}\BibitemShut {NoStop}%
\bibitem [{\citenamefont {Ashkin}(1970)}]{Ashkin_70}%
  \BibitemOpen
  \bibfield  {author} {\bibinfo {author} {\bibfnamefont {A.}~\bibnamefont
  {Ashkin}},\ }\href@noop {} {\bibfield  {journal} {\bibinfo  {journal} {Phys.
  Rev. Lett.}\ }\textbf {\bibinfo {volume} {24}},\ \bibinfo {pages} {156}
  (\bibinfo {year} {1970})}\BibitemShut {NoStop}%
\bibitem [{\citenamefont {Ashkin}\ \emph {et~al.}(1986)\citenamefont {Ashkin},
  \citenamefont {Dziedzic}, \citenamefont {Bjorkholm},\ and\ \citenamefont
  {Chu}}]{Ashkin_86}%
  \BibitemOpen
  \bibfield  {author} {\bibinfo {author} {\bibfnamefont {A.}~\bibnamefont
  {Ashkin}}, \bibinfo {author} {\bibfnamefont {J.~M.}\ \bibnamefont
  {Dziedzic}}, \bibinfo {author} {\bibfnamefont {J.~E.}\ \bibnamefont
  {Bjorkholm}}, \ and\ \bibinfo {author} {\bibfnamefont {S.}~\bibnamefont
  {Chu}},\ }\href@noop {} {\bibfield  {journal} {\bibinfo  {journal} {Opt.
  Lett.}\ }\textbf {\bibinfo {volume} {11}},\ \bibinfo {pages} {288} (\bibinfo
  {year} {1986})}\BibitemShut {NoStop}%
\bibitem [{\citenamefont {Whitesides}(2006)}]{Whitesides2006}%
  \BibitemOpen
  \bibfield  {author} {\bibinfo {author} {\bibfnamefont {G.~M.}\ \bibnamefont
  {Whitesides}},\ }\href@noop {} {\bibfield  {journal} {\bibinfo  {journal}
  {Nature}\ }\textbf {\bibinfo {volume} {442}},\ \bibinfo {pages} {368}
  (\bibinfo {year} {2006})}\BibitemShut {NoStop}%
\bibitem [{\citenamefont {Turing}(1952)}]{Turing}%
  \BibitemOpen
  \bibfield  {author} {\bibinfo {author} {\bibfnamefont {A.~M.}\ \bibnamefont
  {Turing}},\ }\href {\doibase 10.1098/rstb.1952.0012} {\bibfield  {journal}
  {\bibinfo  {journal} {Philosophical Transactions of the Royal Society of
  London. Series B, Biological Sciences}\ }\textbf {\bibinfo {volume} {237}},\
  \bibinfo {pages} {37} (\bibinfo {year} {1952})}\BibitemShut {NoStop}%
\bibitem [{\citenamefont {Jelezko}\ and\ \citenamefont
  {Wrachtrup}(2006)}]{Wrachtrup06}%
  \BibitemOpen
  \bibfield  {author} {\bibinfo {author} {\bibfnamefont {F.}~\bibnamefont
  {Jelezko}}\ and\ \bibinfo {author} {\bibfnamefont {J.}~\bibnamefont
  {Wrachtrup}},\ }\href@noop {} {\bibfield  {journal} {\bibinfo  {journal}
  {Physica status solidi (a)}\ }\textbf {\bibinfo {volume} {203}},\ \bibinfo
  {pages} {3207} (\bibinfo {year} {2006})}\BibitemShut {NoStop}%
\bibitem [{\citenamefont {Balasubramanian}\ \emph {et~al.}(2008)\citenamefont
  {Balasubramanian}, \citenamefont {Chan}, \citenamefont {Kolesov},
  \citenamefont {Al-Hmoud}, \citenamefont {Tisler}, \citenamefont {Shin},
  \citenamefont {Kim}, \citenamefont {Wojcik}, \citenamefont {Hemmer},
  \citenamefont {Krueger}, \citenamefont {Hanke}, \citenamefont
  {Leitenstorfer}, \citenamefont {Bratschitsch}, \citenamefont {Jelezko},\ and\
  \citenamefont {Wrachtrup}}]{Wrachtrup08}%
  \BibitemOpen
  \bibfield  {author} {\bibinfo {author} {\bibfnamefont {G.}~\bibnamefont
  {Balasubramanian}}, \bibinfo {author} {\bibfnamefont {I.~Y.}\ \bibnamefont
  {Chan}}, \bibinfo {author} {\bibfnamefont {R.}~\bibnamefont {Kolesov}},
  \bibinfo {author} {\bibfnamefont {M.}~\bibnamefont {Al-Hmoud}}, \bibinfo
  {author} {\bibfnamefont {J.}~\bibnamefont {Tisler}}, \bibinfo {author}
  {\bibfnamefont {C.}~\bibnamefont {Shin}}, \bibinfo {author} {\bibfnamefont
  {C.}~\bibnamefont {Kim}}, \bibinfo {author} {\bibfnamefont {A.}~\bibnamefont
  {Wojcik}}, \bibinfo {author} {\bibfnamefont {P.~R.}\ \bibnamefont {Hemmer}},
  \bibinfo {author} {\bibfnamefont {A.}~\bibnamefont {Krueger}}, \bibinfo
  {author} {\bibfnamefont {T.}~\bibnamefont {Hanke}}, \bibinfo {author}
  {\bibfnamefont {A.}~\bibnamefont {Leitenstorfer}}, \bibinfo {author}
  {\bibfnamefont {R.}~\bibnamefont {Bratschitsch}}, \bibinfo {author}
  {\bibfnamefont {F.}~\bibnamefont {Jelezko}}, \ and\ \bibinfo {author}
  {\bibnamefont {Wrachtrup}},\ }\href@noop {} {\bibfield  {journal} {\bibinfo
  {journal} {Nature}\ }\textbf {\bibinfo {volume} {455}},\ \bibinfo {pages}
  {648} (\bibinfo {year} {2008})}\BibitemShut {NoStop}%
\bibitem [{\citenamefont {Maze}\ \emph {et~al.}(2008)\citenamefont {Maze},
  \citenamefont {Stanwix}, \citenamefont {Hodges}, \citenamefont {Hong},
  \citenamefont {Taylor}, \citenamefont {Cappellaro}, \citenamefont {Jiang},
  \citenamefont {Gurudev~Dutt}, \citenamefont {Togan}, \citenamefont {Zibrov},
  \citenamefont {Yacoby}, \citenamefont {Walsworth},\ and\ \citenamefont
  {Lukin}}]{Lukin08}%
  \BibitemOpen
  \bibfield  {author} {\bibinfo {author} {\bibfnamefont {J.~R.}\ \bibnamefont
  {Maze}}, \bibinfo {author} {\bibfnamefont {P.~L.}\ \bibnamefont {Stanwix}},
  \bibinfo {author} {\bibfnamefont {J.~S.}\ \bibnamefont {Hodges}}, \bibinfo
  {author} {\bibfnamefont {S.}~\bibnamefont {Hong}}, \bibinfo {author}
  {\bibfnamefont {J.~M.}\ \bibnamefont {Taylor}}, \bibinfo {author}
  {\bibfnamefont {P.}~\bibnamefont {Cappellaro}}, \bibinfo {author}
  {\bibfnamefont {L.}~\bibnamefont {Jiang}}, \bibinfo {author} {\bibfnamefont
  {M.~V.}\ \bibnamefont {Gurudev~Dutt}}, \bibinfo {author} {\bibfnamefont
  {E.}~\bibnamefont {Togan}}, \bibinfo {author} {\bibfnamefont {A.~S.}\
  \bibnamefont {Zibrov}}, \bibinfo {author} {\bibfnamefont {A.}~\bibnamefont
  {Yacoby}}, \bibinfo {author} {\bibfnamefont {R.~L.}\ \bibnamefont
  {Walsworth}}, \ and\ \bibinfo {author} {\bibfnamefont {M.~D.}\ \bibnamefont
  {Lukin}},\ }\href@noop {} {\bibfield  {journal} {\bibinfo  {journal}
  {Nature}\ }\textbf {\bibinfo {volume} {455}},\ \bibinfo {pages} {644}
  (\bibinfo {year} {2008})}\BibitemShut {NoStop}%
\bibitem [{\citenamefont {Yip}\ \emph {et~al.}(2019)\citenamefont {Yip},
  \citenamefont {Ho}, \citenamefont {Yu}, \citenamefont {Chen}, \citenamefont
  {Zhang}, \citenamefont {Kasahara}, \citenamefont {Mizukami}, \citenamefont
  {Shibauchi}, \citenamefont {Matsuda}, \citenamefont {Goh},\ and\
  \citenamefont {Yang}}]{Yang_2019}%
  \BibitemOpen
  \bibfield  {author} {\bibinfo {author} {\bibfnamefont {K.~Y.}\ \bibnamefont
  {Yip}}, \bibinfo {author} {\bibfnamefont {K.~O.}\ \bibnamefont {Ho}},
  \bibinfo {author} {\bibfnamefont {K.~Y.}\ \bibnamefont {Yu}}, \bibinfo
  {author} {\bibfnamefont {Y.}~\bibnamefont {Chen}}, \bibinfo {author}
  {\bibfnamefont {W.}~\bibnamefont {Zhang}}, \bibinfo {author} {\bibfnamefont
  {S.}~\bibnamefont {Kasahara}}, \bibinfo {author} {\bibfnamefont
  {Y.}~\bibnamefont {Mizukami}}, \bibinfo {author} {\bibfnamefont
  {T.}~\bibnamefont {Shibauchi}}, \bibinfo {author} {\bibfnamefont
  {Y.}~\bibnamefont {Matsuda}}, \bibinfo {author} {\bibfnamefont {S.~K.}\
  \bibnamefont {Goh}}, \ and\ \bibinfo {author} {\bibfnamefont
  {S.}~\bibnamefont {Yang}},\ }\href {\doibase 10.1126/science.aaw4278}
  {\bibfield  {journal} {\bibinfo  {journal} {Science}\ }\textbf {\bibinfo
  {volume} {366}},\ \bibinfo {pages} {1355} (\bibinfo {year}
  {2019})}\BibitemShut {NoStop}%
\bibitem [{\citenamefont {Gates}(2009)}]{Gates}%
  \BibitemOpen
  \bibfield  {author} {\bibinfo {author} {\bibfnamefont {B.~D.}\ \bibnamefont
  {Gates}},\ }\href {\doibase 10.1126/science.1171230} {\bibfield  {journal}
  {\bibinfo  {journal} {Science}\ }\textbf {\bibinfo {volume} {323}},\ \bibinfo
  {pages} {1566} (\bibinfo {year} {2009})}\BibitemShut {NoStop}%
\bibitem [{\citenamefont {Kim}\ \emph {et~al.}(2019)\citenamefont {Kim},
  \citenamefont {Campbell}, \citenamefont {de~{\'A}vila},\ and\ \citenamefont
  {Wang}}]{Kim2019}%
  \BibitemOpen
  \bibfield  {author} {\bibinfo {author} {\bibfnamefont {J.}~\bibnamefont
  {Kim}}, \bibinfo {author} {\bibfnamefont {A.~S.}\ \bibnamefont {Campbell}},
  \bibinfo {author} {\bibfnamefont {B.~E.-F.}\ \bibnamefont {de~{\'A}vila}}, \
  and\ \bibinfo {author} {\bibfnamefont {J.}~\bibnamefont {Wang}},\ }\href@noop
  {} {\bibfield  {journal} {\bibinfo  {journal} {Nature Biotechnology}\
  }\textbf {\bibinfo {volume} {37}},\ \bibinfo {pages} {389} (\bibinfo {year}
  {2019})}\BibitemShut {NoStop}%
\bibitem [{\citenamefont {Saggio}\ \emph {et~al.}(2015)\citenamefont {Saggio},
  \citenamefont {Riillo}, \citenamefont {Sbernini},\ and\ \citenamefont
  {Quitadamo}}]{Saggio_2015}%
  \BibitemOpen
  \bibfield  {author} {\bibinfo {author} {\bibfnamefont {G.}~\bibnamefont
  {Saggio}}, \bibinfo {author} {\bibfnamefont {F.}~\bibnamefont {Riillo}},
  \bibinfo {author} {\bibfnamefont {L.}~\bibnamefont {Sbernini}}, \ and\
  \bibinfo {author} {\bibfnamefont {L.~R.}\ \bibnamefont {Quitadamo}},\
  }\href@noop {} {\bibfield  {journal} {\bibinfo  {journal} {Smart Materials
  and Structures}\ }\textbf {\bibinfo {volume} {25}},\ \bibinfo {pages}
  {013001} (\bibinfo {year} {2015})}\BibitemShut {NoStop}%
\bibitem [{\citenamefont {Walker}(2012)}]{Walker_12}%
  \BibitemOpen
  \bibfield  {author} {\bibinfo {author} {\bibfnamefont {G.}~\bibnamefont
  {Walker}},\ }\href {\doibase 10.1002/jsid.100} {\bibfield  {journal}
  {\bibinfo  {journal} {Journal of the Society for Information Display}\
  }\textbf {\bibinfo {volume} {20}},\ \bibinfo {pages} {413} (\bibinfo {year}
  {2012})}\BibitemShut {NoStop}%
\bibitem [{\citenamefont {Awschalom}\ \emph {et~al.}(2018)\citenamefont
  {Awschalom}, \citenamefont {Hanson}, \citenamefont {Wrachtrup},\ and\
  \citenamefont {Zhou}}]{Awschalom_18}%
  \BibitemOpen
  \bibfield  {author} {\bibinfo {author} {\bibfnamefont {D.~D.}\ \bibnamefont
  {Awschalom}}, \bibinfo {author} {\bibfnamefont {R.}~\bibnamefont {Hanson}},
  \bibinfo {author} {\bibfnamefont {J.}~\bibnamefont {Wrachtrup}}, \ and\
  \bibinfo {author} {\bibfnamefont {B.~B.}\ \bibnamefont {Zhou}},\ }\href@noop
  {} {\bibfield  {journal} {\bibinfo  {journal} {Nature Photonics}\ }\textbf
  {\bibinfo {volume} {12}},\ \bibinfo {pages} {516} (\bibinfo {year}
  {2018})}\BibitemShut {NoStop}%
\bibitem [{\citenamefont {Fuchs}\ \emph {et~al.}(2009)\citenamefont {Fuchs},
  \citenamefont {Dobrovitski}, \citenamefont {Toyli}, \citenamefont
  {Heremans},\ and\ \citenamefont {Awschalom}}]{Fuchs_09}%
  \BibitemOpen
  \bibfield  {author} {\bibinfo {author} {\bibfnamefont {G.~D.}\ \bibnamefont
  {Fuchs}}, \bibinfo {author} {\bibfnamefont {V.~V.}\ \bibnamefont
  {Dobrovitski}}, \bibinfo {author} {\bibfnamefont {D.~M.}\ \bibnamefont
  {Toyli}}, \bibinfo {author} {\bibfnamefont {F.~J.}\ \bibnamefont {Heremans}},
  \ and\ \bibinfo {author} {\bibfnamefont {D.~D.}\ \bibnamefont {Awschalom}},\
  }\href {\doibase 10.1126/science.1181193} {\bibfield  {journal} {\bibinfo
  {journal} {Science}\ }\textbf {\bibinfo {volume} {326}},\ \bibinfo {pages}
  {1520} (\bibinfo {year} {2009})}\BibitemShut {NoStop}%
\bibitem [{\citenamefont {Schaibley}\ \emph {et~al.}(2016)\citenamefont
  {Schaibley}, \citenamefont {Yu}, \citenamefont {Clark}, \citenamefont
  {Rivera}, \citenamefont {Ross}, \citenamefont {Seyler}, \citenamefont {Yao},\
  and\ \citenamefont {Xu}}]{Yao_16}%
  \BibitemOpen
  \bibfield  {author} {\bibinfo {author} {\bibfnamefont {J.~R.}\ \bibnamefont
  {Schaibley}}, \bibinfo {author} {\bibfnamefont {H.}~\bibnamefont {Yu}},
  \bibinfo {author} {\bibfnamefont {G.}~\bibnamefont {Clark}}, \bibinfo
  {author} {\bibfnamefont {P.}~\bibnamefont {Rivera}}, \bibinfo {author}
  {\bibfnamefont {J.~S.}\ \bibnamefont {Ross}}, \bibinfo {author}
  {\bibfnamefont {K.~L.}\ \bibnamefont {Seyler}}, \bibinfo {author}
  {\bibfnamefont {W.}~\bibnamefont {Yao}}, \ and\ \bibinfo {author}
  {\bibfnamefont {X.}~\bibnamefont {Xu}},\ }\href@noop {} {\bibfield  {journal}
  {\bibinfo  {journal} {Nature Reviews Materials}\ }\textbf {\bibinfo {volume}
  {1}},\ \bibinfo {pages} {16055} (\bibinfo {year} {2016})}\BibitemShut
  {NoStop}%
\bibitem [{\citenamefont {Park}\ \emph {et~al.}(2013)\citenamefont {Park},
  \citenamefont {Vosguerichian},\ and\ \citenamefont {Bao}}]{Bao_2013}%
  \BibitemOpen
  \bibfield  {author} {\bibinfo {author} {\bibfnamefont {S.}~\bibnamefont
  {Park}}, \bibinfo {author} {\bibfnamefont {M.}~\bibnamefont {Vosguerichian}},
  \ and\ \bibinfo {author} {\bibfnamefont {Z.}~\bibnamefont {Bao}},\
  }\href@noop {} {\bibfield  {journal} {\bibinfo  {journal} {Nanoscale}\
  }\textbf {\bibinfo {volume} {5}},\ \bibinfo {pages} {1727} (\bibinfo {year}
  {2013})}\BibitemShut {NoStop}%
\bibitem [{\citenamefont {High}\ \emph {et~al.}(2015)\citenamefont {High},
  \citenamefont {Devlin}, \citenamefont {Dibos}, \citenamefont {Polking},
  \citenamefont {Wild}, \citenamefont {Perczel}, \citenamefont {de~Leon},
  \citenamefont {Lukin},\ and\ \citenamefont {Park}}]{High2015}%
  \BibitemOpen
  \bibfield  {author} {\bibinfo {author} {\bibfnamefont {A.~A.}\ \bibnamefont
  {High}}, \bibinfo {author} {\bibfnamefont {R.~C.}\ \bibnamefont {Devlin}},
  \bibinfo {author} {\bibfnamefont {A.}~\bibnamefont {Dibos}}, \bibinfo
  {author} {\bibfnamefont {M.}~\bibnamefont {Polking}}, \bibinfo {author}
  {\bibfnamefont {D.~S.}\ \bibnamefont {Wild}}, \bibinfo {author}
  {\bibfnamefont {J.}~\bibnamefont {Perczel}}, \bibinfo {author} {\bibfnamefont
  {N.~P.}\ \bibnamefont {de~Leon}}, \bibinfo {author} {\bibfnamefont {M.~D.}\
  \bibnamefont {Lukin}}, \ and\ \bibinfo {author} {\bibfnamefont
  {H.}~\bibnamefont {Park}},\ }\href@noop {} {\bibfield  {journal} {\bibinfo
  {journal} {Nature}\ }\textbf {\bibinfo {volume} {522}},\ \bibinfo {pages}
  {192} (\bibinfo {year} {2015})}\BibitemShut {NoStop}%
\end{thebibliography}%
\textbf{Acknowledgments} We thank R. Kolesov, J. Hofkens, M. Totzeck, J. Wrachtrup and W. Yao for the fruitful discussions. We thank F. Sardi, T. Shalomayeva, S.K. Li, C.H. Lai and M.H. Yeung for the technical helps. S.Y. acknowledges financial support from CUHK start-up grant.
\\
\\
\\
\textbf{Materials \& Correspondence} Correspondence and requests for materials should be addressed to K.X.~(email: kangweixia@gmail.com) or S.Y.~(email: syang@cuhk.edu.hk).

\end{document}